# Your Neighbors Are My Spies: Location and other Privacy Concerns in Dating Apps

Nguyen Phong HOANG, Yasuhito ASANO, Masatoshi YOSHIKAWA

Department of Social Informatics, Graduate School of Informatics, Kyoto University, Japan

hoang.nguyenphong.jp@ieee.org, asano@i.kyoto-u.ac.jp, yoshikawa@i.kyoto-u.ac.jp

*Abstract*— **Trilateration has recently become one of the well-known threat models to the user's location privacy in location-based applications (aka: location-based services or LBS), especially those containing highly sensitive information such as dating applications. The threat model mainly depends on the distance shown from the targeted victim to the adversary to pinpoint the victim's position. As a countermeasure, most of location-based applications have already implemented the "hide distance" function to protect their user's location privacy. The effectiveness of such approach however is still questionable. Therefore, in this paper, we first investigate how popular location-based dating applications are currently protecting their user's privacy by testing the two most popular GLBT-focused applications: Jack'd and Grindr. As one of our findings, we then demonstrate how an adversary can still figure out the location of the targeted victim even when the "hide distance" function is enabled. Our threat model is simply an enhanced version of the trilateration model. Without using sophisticated hacking tools or complex attack techniques, the model is still very effective and efficient at locating the targeted victim, and of course in a so-called "legal" manner since we only utilize the information that can be obtained just as same as any other ordinary user. In addition, we also introduce a potential side channel attack fashion due to the current design of Jack'd. Our study thus raises an urgent alarm to those location-based applications' users in general, and especially to those GLBT-focused dating application's users about their privacy. Finally, the paper concludes by suggesting some possible solutions from the viewpoints of both the provider and the user considering the implementation cost and the trade-off of utility.**

## I. Introduction

Nowadays, thanks to the advancement of Global Positioning System (GPS), most of current smart phones have a built-in GPS receiver, which helps to estimate the location information with accuracy up to just a few meters. Taking this advantage, location-based applications are getting more dominant in the smart phone applications market. Just a decade ago, one still had to use paper map, or ask for direction when going to an unknown area; while young people were surfing around online chat rooms to look for friends at that time. However, the introduction of LBSs has changed our lifestyle and the way that people interact with each other thanks to its undeniable convenience. For instance, one can easily find the nearest restaurant, convenience store or shopping mall by using application like Google Map; or hang out with friends by using application like Find My Friends, etc.

### A. Privacy in General

Nevertheless, in the era of Information and Communication Technology, along with censorship and massive surveillance in cyberspace, the problem of information leakage has also become more and more severe. Tim Cook, the CEO of Apple Inc., used to say at the White House Cyber Security Summit in early 2015 that: "Privacy is a matter of life and death". As people increasingly keep more sensitive personal information in their phone, big agencies and companies like Apple have been working hard to provide the best protection to their customer's private information. However, an absolute privacy and a completely perfect countermeasure for preventing future data breaches still remain as headache matters. According to the Tenth Annual Cost of Data Breach Study published by IBM in 2015, the average consolidated total cost of a data breach is $3.8 million, increasing 23% since 2013. The report also points out that the cost incurred for each lost or stolen record that has sensitive and confidential information increased 6% from a consolidated average of $145 to $154 [1].

Among personally identifiable information, location is considered as one of the most essential factors since the leak of location information can later lead to the disclosure of other sensitive private information such as occupation, hobbies, daily routines, social relationships [14]. In spite of many attack techniques [5], [11] that have been studied by the research community since then, the protection of location privacy from both LBS provider side and user side has not been sufficiently and appropriately taken into account. Thus, in this paper, we investigate the current status of location privacy preserving in popular GLBT-focused dating apps to have a clearer view, and observe how it is being protected in real-life practice with both already-known threat model and its enhanced version.

### B. Privacy Concerns in GLBT-focused Apps

It is important to emphasize that it is not because of hatred or discrimination that makes us opt for investigating GLBT-focused applications like Jack'd and Grindr. But, because of their popularity, possession of highly-sensitive information, and the huge number of users[1] that make the apps highly vulnerable to cyber-attack just as same as the case of Ashley Madison [3]. In addition, it is also because GLBT-focused dating applications like Grindr and Jack'd are location sensitive, and their users depend on the publicly shown distance information to look for nearby people to meet up right away, thus potentially exposed to the risk of being located.

As stated in [12], there are still many Islamic nations where homosexuality carries the death penalty. For instance, there were several gay men in Syria lured by ISIS terrorists to go out on dates, and later executed publicly by stoning as recently reported in [4]. Even in those regions like North America and Western Countries, which are thought to be more open-minded, the GLBT community is still not widely accepted. More or less, people belong to the GLBT community are still facing the problem of being attacked, harassed or discriminated. Such cases show that protecting privacy of GLBT-focused application's user is a nontrivial task, and should not be neglected by the LBS providers. Because the location information together with other information such as height, weight, age, hobbies can be employed to accurately disclose the targeted individuals. Later, the compromised information from those victims such as frequently visiting place, occupation, address, daily routines and social relationships can be used to intimidate for money, or even lead to physical harassment. At this point, things become understandable why Tim Cook says: "Privacy is a matter of life and death" since he also came out as gay in October 2014.

### C. Organization

The rest of this paper is organized as follows. We will introduce our experimental environment in the last part of this section. In Section II, Grindr and Jack'd are investigated in terms of location privacy. We will demonstrate how the user's location can be accurately discovered even when approaches like location anonymization have been implemented. In Section III, other privacy concerns are discussed. Additionally, we introduce a side channel attack fashion that can be conducted due to current design of Jack'd. Finally, from the viewpoints of both LBS provider and user, we then give some possible solutions, and conclude the paper in Section IV.

### D. Experiment Setup

The trilateration threat model actually can be conducted in a physical way that the adversary carries his device around to three different places and notes down the distances shown from himself to the victim. However, in order to have an easily manageable experimental environment, we employ three virtual machines that host Android OS to play the role of adversaries. Each machine is then set to be in positions as following:

---

[1] According to Grindr, the app has around two million daily active users [8], while Jack'd has 5 million users as of 2014 with more than two million active users [10].

- Victim is an account run on a real iPhone 5, locates at Science Frontier Laboratory, Kyoto University with coordinates (35.02350485, 135.77687703).
- Adversary A1 is located at Demachi-yanagi Station with coordinates (35.03051251, 135.77327415).
- Adversary A2 is located at Heian Shrine with coordinates (35.01598257, 135.78242585).
- Adversary A3 is located at Kyoto Imperial Palace with coordinates (35.02258561, 135.76493382).

Each Android machine is then equipped with Fake-GPS so that their positions can be freely set to any corner of the world. Next, to capture packets in Subsection III.A, we use Microsoft Network Monitor version 3.4. All of the maps used in this study are drawn by using a free map tool available at: http://obeattie.github.io/gmaps-radius/.

## II. LOCATION PRIVACY CONCERN

### A. Trilateration Model

To the best of our knowledge, the trilateration threat model is allegedly said to be first reported to Grindr in 2014 [9], and discussed in recent studies [5] and [11]. The main idea of this attack model bases on the distance from the user, which is publicly shown to other users. With privilege no more than an ordinary user, an adversary just needs to move around the victim to three different places. Distances from the victim to the adversary at those three positions are then used to pinpoint the exact location of the victim. As shown in Figure 1, Grindr user, who opts for showing his distance, is facing high risk of being located since the adversary can obtain accuracy up to the victim's building as highlighted in the red rectangle. This threat model works totally the same with Jack'd.

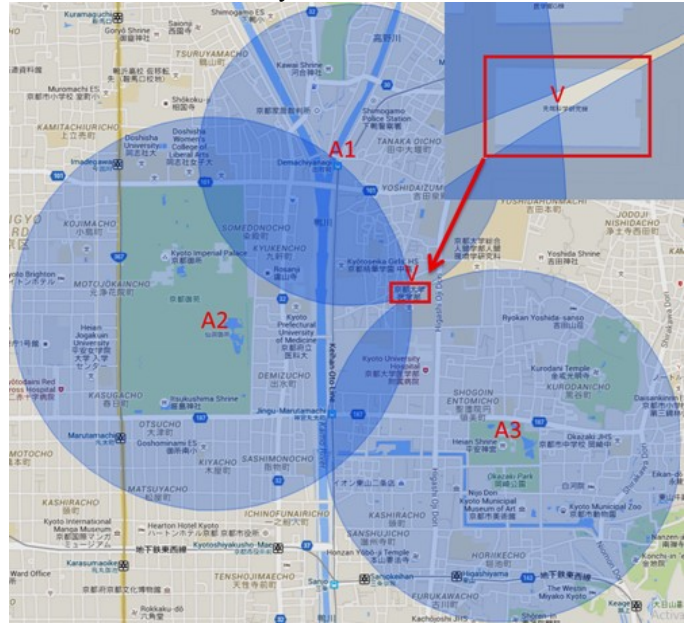

**Figure 1.** Testing Trilateration Threat Model in Grindr

From the geometry point of view, the location of the victim is nothing else but the coordinates of V, which is the solution $(x, y)$ of a system of simultaneous circle equations.

$$\begin{cases}(x - x_{A1})^2 + (y - y_{A1})^2 = D1^2 \\ (x - x_{A2})^2 + (y - y_{A2})^2 = D2^2 \\ (x - x_{A3})^2 + (y - y_{A3})^2 = D3^2\end{cases}$$

Where:

- $(x_{A1}, y_{A1}), (x_{A2}, y_{A2}), (x_{A3}, y_{A3})$ are latitudes and longitudes of the adversary at three different positions $A1, A2, A3$ respectively.
- $D1, D2, D3$ are distances from V to $A1, A2, A3$ respectively.

In response to this type of attack, Grindr adopted a function in which the user can opt to hide the distance since August 2014 [7]. Thus, the threat of trilateration model is disabled to locate those users who already enabled the "hide distance" function. As we revisited [9] at the time of writing this paper, the map was no longer able to pinpoint Grindr users as shown in Figure 2.

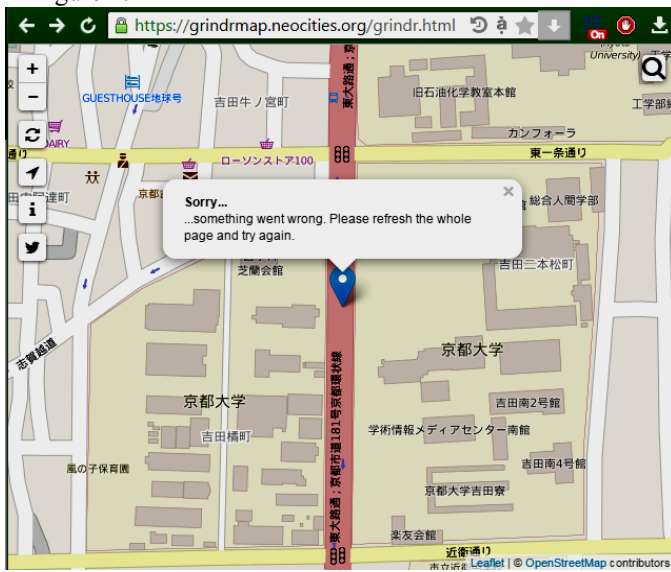

**Figure 2.** Previous Grindr's flaw had been fixed.

### B. Your Neighbors are My Spies

Despite of the fact that one of the best solutions to protect the location information is not to publish it; in this part, as a key point of this paper, we will illustrate an enhanced version of the trilateration threat model that current approach like location anonymization implemented by enabling the "hide distance" function still cannot effectively counter to. The primary factor in success of this threat is based on the way that Grindr arranges its users on the screen. Perhaps, in order to maintain the utility, users are displayed left-to-right and top-to-down in an ascending order of their distances regardless of whether they have disabled the "show distance" function or not. By exploiting this fact, the neighbors displayed just before and just behind the victim unintentionally become the upper and lower bounds of the distance from the victim to the adversary. As a result, the region in which the victim is locating is easily obtained by employing the trilateration model again, but with two circles drawing from the adversary to the two nearest neighbors as shown in Figure 3.

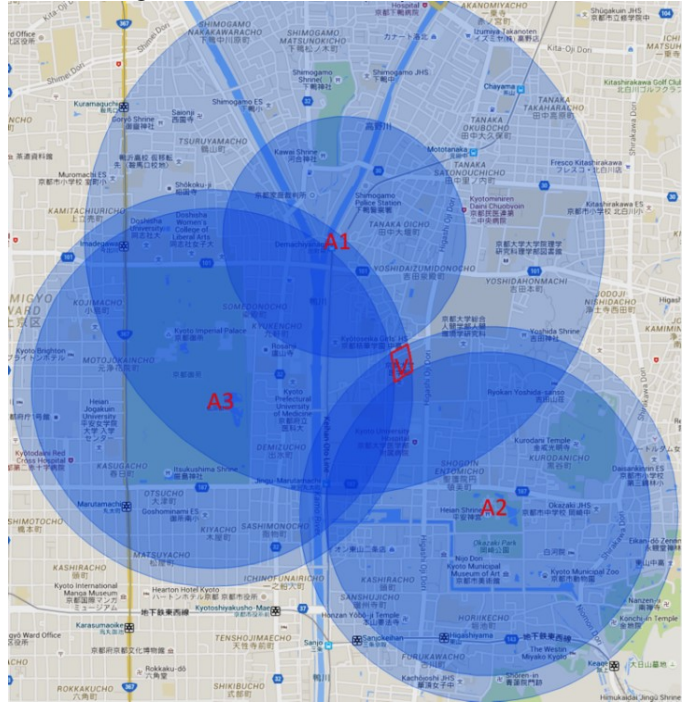

**Figure 3.** Enhanced version of Trilateration Threat Model

By using this threat model, the adversary even does not need to query the victim profile several times to note down the distance as done with the original trilateration model, but still very effective at locating the victim's region as marked in Figure 4.

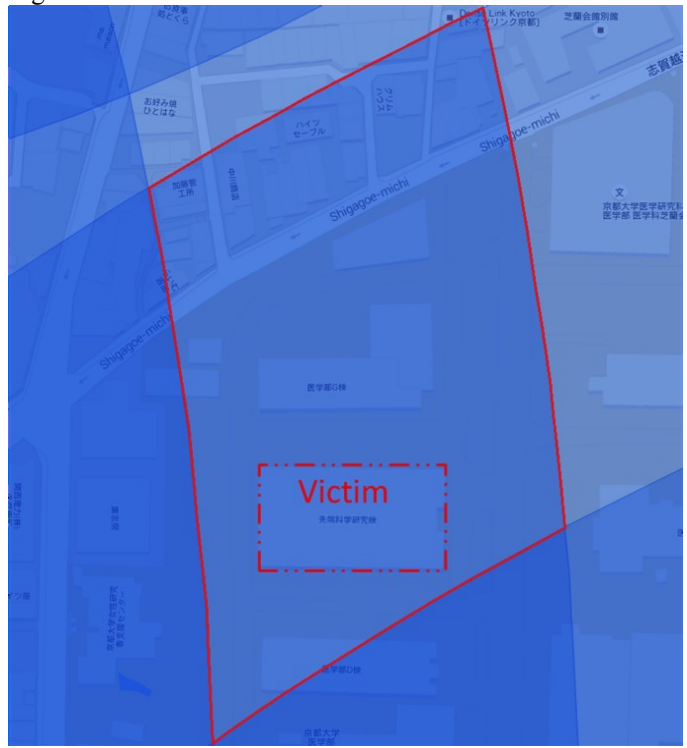

**Figure 4.** The victim's region is smoothly bounded.

A lesson learned from Figure 3 and Figure 4 is that the adversary gains more location information about the victim at A2 and A3 than A1 because the bounds set by the pairs of victim's neighbors from the viewpoints of A2 and A3 are more narrow than from A1. Therefore, even without drawing the pair circles from A1, the adversary can still confidently infer the possible region of the victim, because one of the two intersection areas is a river space, thus the probability that the victim is locating in that region is relatively low.

To this point, one may think that this type of attack model will not work with those victims who locate in low-density area since there are not so many neighbors around him for this attack model to take place. However, as far as we are concerned, using the term "victim's neighbors" actually is not always correct, because physically they may not locate near to the victim. In fact, this attack model is still valid as long as the following condition holds:

$$AN1 < AV < AN2$$

Where:
- AV is distance from the adversary to the victim.
- AN1 and AN2 are distances from the adversary to the pair of so-called victim's nearest neighbors.

Thus, in real life attack, the adversary can also apply this model to attack the victim in remote area by placing himself in the middle of high-density areas and the victim's region as shown in Figure 5. The more crowded the urban areas are, the more resources that the adversary can possess to precisely explore the victim's location. Or in a more active attack fashion, the adversary can create two colluding accounts and move them around until he can satisfactorily compromise the victim's location. The key idea is to gradually reduce the subtraction value of |AN1-AN2| such that V is still sandwiched by N1 and N2 on the display screen of the adversary. The smaller the subtraction value becomes, the more accurate location of the victim can be revealed.

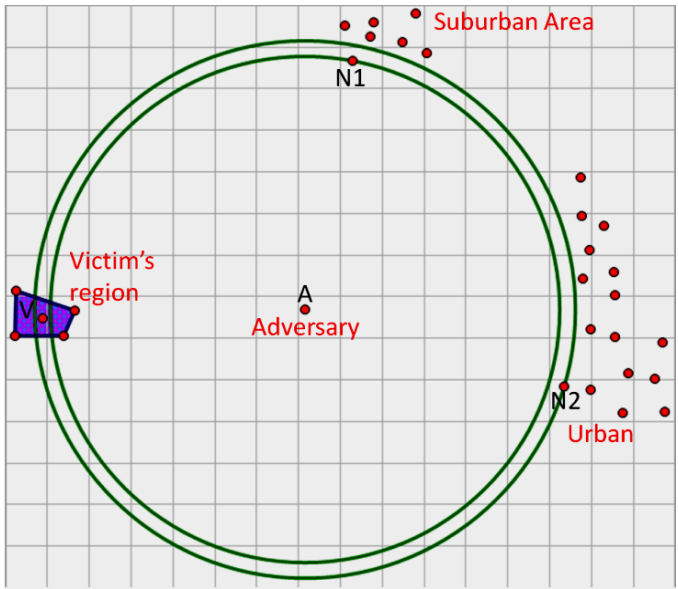

**Figure 5.** The "Victim's neighbors" are not necessarily close to him.

Up to this point, it is obvious that obtaining victim's real location from Figure 4 becomes a trivial task. Moreover, with this enhanced model, even when all local members hide their distances, the adversary can still make complete use of his colluding accounts to infer, thus be able to narrow down the possible region in which the victim is locating.

### III. OTHER PRIVACY CONCERNS

#### A. Vulnerabilities from Third-Party Advertisement Banner and Misconduct in Handling User's Personal Information

In the age of online marketing, one's privacy is often threatened by the very advertisements popping up in his device as mentioned in [2]. In order to investigate this issue in Jack'd and Grindr, we analyze packets captured while using these two applications. The results are shown in Figure 6.

```
229..x-failurl: http://ads.mopub.com/m/ad?v=8&udi
d=ifa:E121C9CB-9758-4076-9DE1-422BB33F3F84&id=agl
tb3B1Yi1pbmNyDQsSBFNpdGUY7cz7Bgw&nv=1.17.2.0&q=m_
gender:m,m_age:0&o=p&sc=2.0&z=+0900&mr=1&ct=2&av=
2.2.4&cn=KDDI&iso=jp&mnc=50&mcc=440&dn=iPhone5%2C
2&exclude=46a0e29f574448038760b6e06a3232f4&reques
t_id=e7918915a00144508c4afd5fa7fcbb91&fail=1..x-i
mptracker: http://ads.mopub.com/m/imp?appid=&cid=
31fd6d9c46d14787b072ca69dee8b44c&city=&ckv=2&coun
try_code=JP&cppck=2375D&dev=iPhone5%2C2&id=agltb3
B1Yi1pbmNyDQsSBFNpdGUY7cz7Bgw&is_mraid=1&mpx_clk=
http%3A%2F%2Fmpx.mopub.com%2Fclick%3Fad_domain%3D
agoda.com%26adgroup_id%3D46a0e29f574448038760b6e0
6a3232f4%26adunit_id%3Dagltb3B1Yi1pbmNyDQsSBFNpdG
UY7cz7Bgw%26ads_creative_id%3D31fd6d9c46d14787b07
2ca69dee8b44c%26app_id%3Dagltb3B1Yi1pbmNyCwsSA0Fw
cBiJwSEM%26app_name%3DGrindr%2520iOS%26auction_ti
me%3D1444444014%26bid_price%3D19.49%26bidder_id%3
4..x-failurl: http://ads.mopub.com/m/ad?v=8&udid=
ifa:BBC656C1-3F0B-4B6A-8D85-77D7E1D5476C&id=d7ea3
f8c3825497f940bf56b05335665&nv=3.3.0&o=p&sc=2.0&z
=+0900&ll=35.02353627550613,135.776885205088&lla=
65&llsdk=1&mr=1&ct=2&av=3.1&cn=KDDI&iso=jp&mnc=50
&mcc=440&dn=iPhone5%2C2&ts=1&request_id=cdacc5b2d
47445098c877eb5b4202bd1&fail=1&fail=1&exclude=628
0a0eef1a14dd58f421e5c4cc0a94b&exclude=74de412afe2
611e38aab1231392559e4&exclude=75092462fe2611e38aa
b1231392559e4&fail=1..x-height: 50..x-imptracker:
http://ads.mopub.com/m/imp?appid=&cid=03faca92f1
4c4f6b9d6896069663eb18&city=&ckv=2&country_code=J
P&cppck=FC310&dev=iPhone5%2C2&id=d7ea3f8c3825497f
940bf56b05335665&is_mraid=0&mpx_clk=http%3A%2F%2F
mpx.mopub.com%2Fclick%3Flineitem%3D5217644%26ad_d
omain%3Dwish.com%26ad_id%3D666eba282b3623b1%26adg
roup_id%3D6280a0eef1a14dd58f421e5c4cc0a94b%26adun
it_id%3Dd7ea3f8c3825497f940bf56b05335665%26ads_cr
eative_id%3D03faca92f14c4f6b9d6896069663eb18%26ap
p_id%3D09f13c886c604d25a524584215881989%26app_nam
e%3DJack%25E2%2580%2599d%2520-%2520iOS%26auction_
time%3D1444398521%26bid_price%3D0.13%26bidder_id%
```

**Figure 6.** Information Leakage through Third-Party Ads

Surprisingly, in both applications, the third-party advertisements leak many important information of the users including name of Telecommunications Service Provider (i.e., KDDI), device's model information (i.e., iPhone 5), country code (i.e., JP), and last but not least: the application name (i.e., Grindr, Jack) which is the most sensitive information that no any straight-acting person wants the others to know the existence of such applications in his phone. It may have no problem if the packets are sent directly to the ads provider's server. However, what is worth mentioning here is that the packets are sent in an unencrypted fashion, thus widely open to an attack type known as man-in-the-middle attack, in which the hacker taps the Internet connection of the victim to eavesdrop the packets.

By analyzing all the captured packets, we were further shocked by the fact that all the packets containing members' profile pictures of Grindr are also sent in the air without encryption, thus being captured and converted into the original image files as shown in Figure 7.

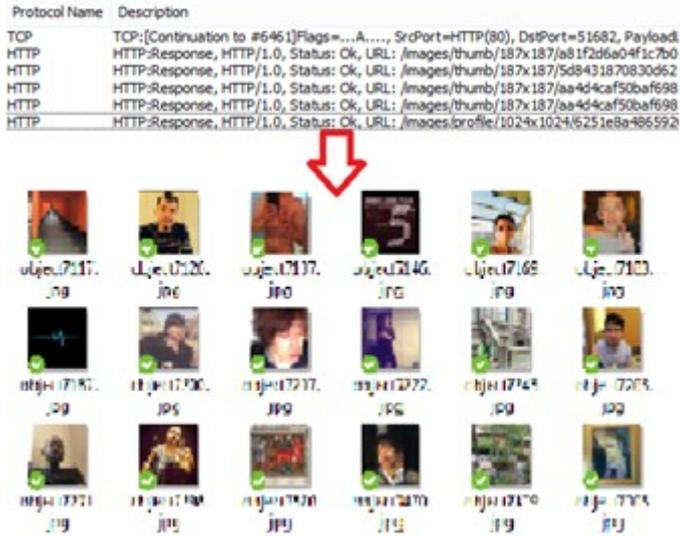

Figure 7. Pictures recovered from unencrypted packets

Concerning ethical issue, we only recover image files that appear on the first screen page of Grindr as evidence, and the users' avatars are intentionally censored. Taking into account of the above findings, it is obvious that the user's privacy is not guaranteed at all regardless of the fact that the vendor has been alerted to these issues by a security firm before [9].

*B. Together with IP Spy and Linkage Attack*

As human being is born curious, social engineering intrusion techniques like phishing is always the easiest but effective way to compromise people. Since Jack'd provides it users with a feature to see who viewed his profile with exact timestamp, an adversary may put an IP-spy URL into his profile to promote his appearance, thus being able to obtain the victim's IP address if the victim feels curious and click on that URL. Nevertheless, it was discussed in [13] that IP address is also important personal information which can be exploited to perform the linkage attack to retrieve other related personal information. To have a clearer view on how this trick is really effective at luring innocent users, we place our Jack'd accounts in three big cities of Japan which are Tokyo, Osaka and Kyoto within 12 hours (from 6PM to 6AM of the following day) to estimate how many innocent and curious victims could be lured. We choose this time period because it is the most active usage time according to [6]. To conduct this task, we need to reboot the virtual machines every 2 hours so that our accounts will not be disappeared from the screen of other local users, because we found that Jack'd only keeps an account displayed on other user's screen for 2 hours since the latest login. The result is illustrated in Figure 8.

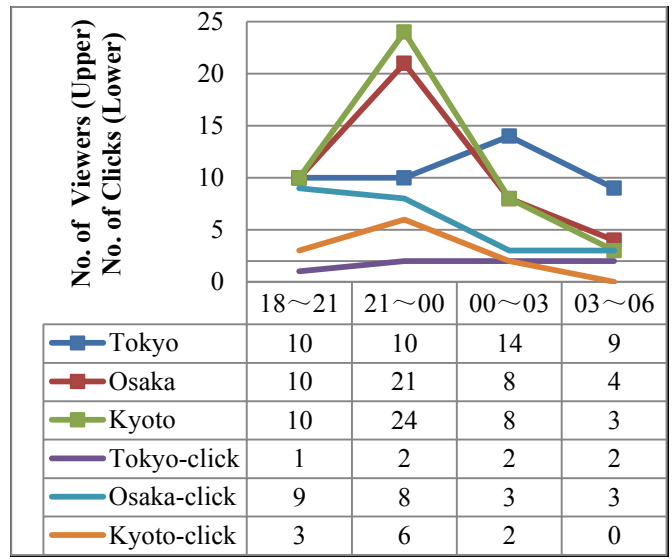

|  | 18～21 | 21～00 | 00～03 | 03～06 |
|---|---|---|---|---|
| Tokyo | 10 | 10 | 14 | 9 |
| Osaka | 10 | 21 | 8 | 4 |
| Kyoto | 10 | 24 | 8 | 3 |
| Tokyo-click | 1 | 2 | 2 | 2 |
| Osaka-click | 9 | 8 | 3 | 3 |
| Kyoto-click | 3 | 6 | 2 | 0 |

Figure 8. Analysis from IP-spy Intrusion Technique

In total, we got 131 viewers from three accounts with 41 viewers clicked on the IP-spy URL that we put in our profile.

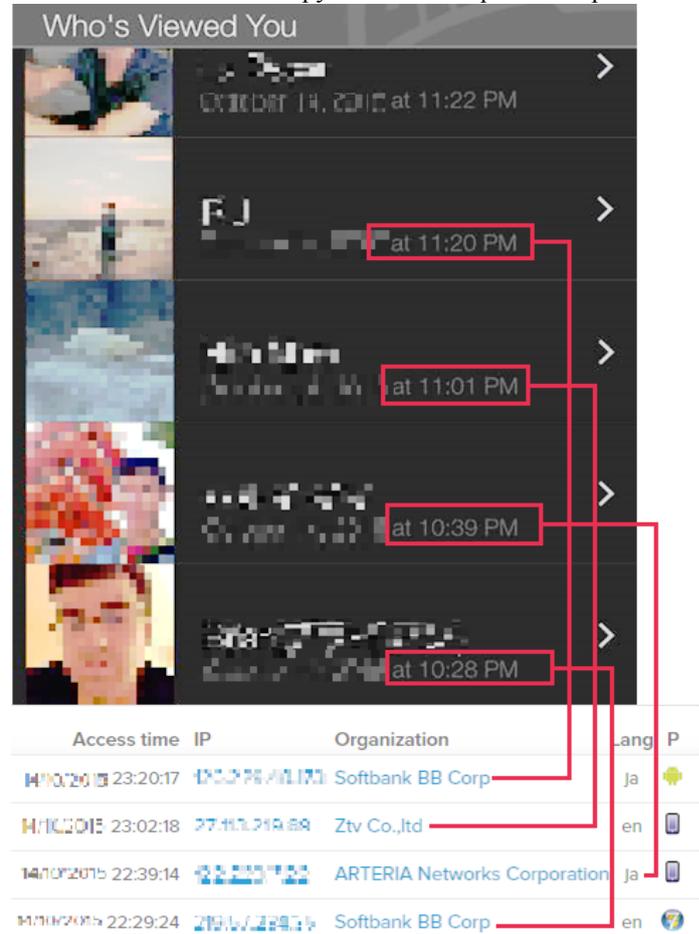

Figure 9. Linkage Attack with IP-spy and Jack'd timestamp

Among these 41 clicks, we were able to perform linkage attack to 26 users with high confidence by matching the timestamp between Jack'd and our IP-spy server to further reveal other information including their IP address, ISP, display language and platform of their devices as shown in Figure 9.

## IV. DISCUSSION AND CONCLUSION

Through this study, we would like to particularly alert the users of Grindr and Jack'd as well as the users of other LBSs in general about the risk of being located easily regardless of whether the recent location anonymization approach has been adopted. By investigating these two apps, we found a paradox that although there have been many attack models proposed by the privacy-preserving researchers, the user's location privacy has not been seriously taken into consideration by both LBS provider and the users themselves. As far as we are concerned, the reason of this negligence comes from both LBS provider and user. From the viewpoint of LBS provider, it might cause overhead to implement those sophisticated solutions proposed by the research community, while the utility of the app is not really guaranteed, thus probably lead to the loss of its customer. From the viewpoint of the user, it may be because of two reasons. Perhaps, the first one is also due to the trade-off of utility. The other one is because of unawareness. For that reason, instead of using complicated mathematical equations and complex algorithms to show the threat model, we opt for visualizing it on map in this paper so that even those non-technical readers can understand how easily their privacy can be compromised in the current security condition.

In order to manage risks of man-in-the-middle attack, IP spy and other side channel attacks as mentioned in Section III, we urge the LBS providers and involved third-parties to carefully encrypt the connection from their servers to the users. For the user side, we suggest not opening any URL out of curiosity. If it is really necessary to open an unknown URL, the user should turn on VPN at first to prevent the leak of their real IP address.

For those threat models discussed in Section II, let us argue that privacy preserving policy is different from person to person. Especially in GLBT community, some already came out, thus have no concern about privacy; while some are straight-acting, thus do not want to be disclosed. Therefore, a centralized solution is not really suitable, and users are the very ones who need to make decision whether to protect their own privacy. For the meantime, while waiting for the experts and the LBS providers to discover a perfect solution for location privacy protection without trading off the utility of LBS, we suggest that the user should take a step forward to protect them from those vulnerabilities mentioned in this study. That is to use exactly the Fake-GPS applications like the one that we use in our experiment (probably also used by most of the adversaries) to hide the real location to an acceptable extent so that the user can still gain the convenience provided by the LBS. How far the fake location should be shifted from the real one depends on how much utility that a user is willing to trade off with his privacy, thus it is different from case by case. We strongly believe that this user-centric solution not only suits all type of users, but also helps to save the vendors from overhead investment in implementing sophisticated solutions and infrastructures.

Apart from aforementioned technical methods, human factor is also important in protecting oneself in the cyberspace. In order to avoid troublesome problems in the future when the vendors get hacked as the case of Ashley Madison [3], the user should not register account to those highly sensitive apps under his real name or even a part of his real name. Instead, the user should use information that could not be used to link the account with his real-life personally identifiable information.

Last but not least, in this paper, we of course could have utilized higher techniques to extract and test the threat model's accuracy with more users of Jack'd and Grindr in bulk. However, as far as we are concerned with the ethical issue that those compromised users also have their right to be undisclosed, and there may be our acquaintances among them. We thus did not go beyond those accounts created by ourselves.

## ACKNOWLEDGMENT

This work was supported by JSPS KAKENHI Grant Number 15K00423 and the Kayamori Foundation of Informational Science Advancement.